\documentclass[twocolumn,showpacs,showkeys,preprintnumbers]{revtex4}%
\usepackage{amssymb}
\usepackage{makeidx}
\usepackage{amsmath}
\usepackage{graphicx}
\usepackage{dcolumn}
\usepackage{bm}
\usepackage{amsfonts}%
\providecommand{\U}[1]{\protect\rule{.1in}{.1in}}

\begin{document}
\title{Effect of electric current on optical response of viscous electron-hole plasma}
\author{Yu. A. Pusep$^{1,\ast}$, M. A. T. Patricio$^{2}$, G. M. Jacobsen$^{2}$, M. D.
Teodoro$^{2}$, G. M. Gusev$^{3}$, and A. K. Bakarov$^{4}$\bigskip}

\begin{abstract}
The influence of the Hall voltage on the photoluminescence of a dense
hydrodynamic electron-hole plasma laser generated in a mesoscopic n-doped GaAs
channel under intense laser excitation is studied. Laser excitation induces an
interband current determined by the recombination of photogenerated
electron-hole pairs. As a result, background electrons drifting under the
influence of the Hall voltage form an effective Hall current. The Coulomb drag
caused by the Hall current causes the accumulation of light holes, leading to
the appearance of a double photoluminescence line formed by the recombination
of excitons and trions. In contrast, in the absence of a Hall current, the
shift in the photoluminescence energy associated with heavy holes occurs due
to the electric field created by the Hall potential difference.

\end{abstract}
\date{08/04/2026}
\affiliation{$^{1}$S\~{a}o Carlos Institute of Physics, University of S\~{a}o Paulo, PO Box
369,13560-970 S\~{a}o Carlos, SP, Brazil.}
\affiliation{$^{2}$Physics Department, Federal University of S\~{a}o Carlos, 13565-905,
S\~{a}o Carlos, SP, Brazil.}
\affiliation{$^{3}$Institute of Physics, University of S\~{a}o Paulo, 135960-170 S\~{a}o
Paulo, SP, Brazil.}
\affiliation{$^{4}$Institute of Semiconductor Physics, 630090 Novosibirsk, Russia}
\affiliation{E-mail: pusep@ifsc.usp.br\footnotetext{* Author to whom any correspondence
should be addressed.}}
\startpage{1}
\endpage{102}

\maketitle

\section{Introduction}

The effects of electric fields on the optical properties of bulk
semiconductors are well known, for example, the Stark effect, field-induced
exciton dissociation, band bending and spatial charge redistribution
\cite{sze}. Another important topic in optoelectronics is the effect of
electric current on the optical properties of heterojunctions, where current
injection can affect charge carrier populations, energy levels, and internal
electric fields, which in turn alter optical properties such as absorption,
emission wavelength, refractive index, and optical gain. At the same time, the
most well-known effect of electric current in bulk semiconductors is Joule
heating \cite{allen2020,emi2021}, which causes a narrowing of the band gap, a
change in the refractive index and, therefore, affects the intensity and
spectral range of the optical response.\cite{william2020}. In general, the
study of electro-optical effects is an active area of {}{}research that
combines solid-state physics, optoelectronics and materials science and thus
determines stability and optical output of optoelectronic devices.

In GaAs, electro-optical effects are highly efficient due to its direct
bandgap and non-centrosymmetric crystal structure. These properties allow the
material to change its refractive index or absorption coefficient in response
to an applied electric field, which is fundamental for high-speed optical
modulators and switches. Among the most well-known and important electro-optic
effects is the Franz-Keldysh effect, which is associated with the process of
"photon-mediated tunneling" that occurs in bulk GaAs when a strong electric
field is applied \cite{pintus2019} and Quantum Confined Stark effect, observed
when an electric field is applied perpendicular to the quantum well (QW),
\cite{bloemer1993}. Both of these effects result in a redshift of the
corresponding optical response.

In this study, we report the effect of electric current on the
photoluminescence (PL) of a hydrodynamic electron-hole (e-h) system caused by
the e-h Coulomb drag. Mutual e-h Coulomb drag is characteristic of
hydrodynamic e-h plasmas, occurring when interparticle momentum-conserving
collisions prevail over momentum-relaxing scattering mechanisms. This
phenomenon is driven by effective frictional forces generated through Coulomb
interactions \cite{fritz2024}\textbf{. }In our latest papers
\cite{pusep2025a,pusep2025b}, we investigated the influence of electric
current on the optical response of hydrodynamic e-h plasma formed in a
mesoscopic GaAs channel. We demonstrated that the scattering between electrons
and holes, dominating over disorder scattering, naturally leads to their
mutual Coulomb drag \cite{pusep2025a}. The e-h drag provides conditions for
increasing the local charge concentration, which in turn promotes the
formation of excitons and their complexes, such as excitons and trions. As a
result, electric current-induced magnetoexcitons and trions were observed
\cite{pusep2025b}. In these experiments, an electric current flowed through
the channel, and a magnetic field helped the formation of excitons and exciton
complexes in dense e-h plasmas by compressing the exciton wave function, which
led to an increase in the exciton binding energy and, consequently, to an
increase in the critical electron density at which excitons dissociate.

Here we report further evidences for the influence of electric current on
optical response of the e-h plasma and demonstrate a clear distinction between
it and the effect of the electric field. In the presented experiments, no
current flows inside the channel. Instead, a DC current flows perpendicular to
the channel through the lateral potentiometric contact pads. In this case, the
magnetic field performed an additional action, providing a Hall voltage across
the channel.

\section{Experimental}

The sample used in this study is a mesoscopic channel 5 $\mu$m wide and 100
$\mu$m long fabricated from a single GaAs quantum well (QW), with a thickness
of 14 nm, grown on a (100)-oriented GaAs substrate by a molecular beam
epitaxy. QW barriers were grown in the form of short-period GaAs/AlAs
superlattices. The desired electron concentration was achieved using Si
$\delta$-doping, separated from the QW by 4.5 nm thick spacers. The sheet
electron density and the mobility measured at the temperature of 1.4 K were
9.1$\cdot$10$^{11}$ cm$^{-2}$ and 2.0$\cdot$10$^{6}$ cm$^{2}$/Vs,
respectively. In this structure, the formation of viscous e-h plasma was shown
in Refs.\cite{pusep2022,pusep2023,pusep2024}.

Scanning magneto-photoluminescence (PL) microscopy measurements were performed
at the temperature 4 K using a helium closed cycle cryostat equipped with a
superconducting magnet (Attocube/Attodry1000). A laser beam with a wavelength
of 440 nm (2.82 eV) generated e-h pairs in a spot of about 1 $%
\mu
$m in size, focused in the middle of the channel. The PL spatial resolution of
the setup is determined by the size of the light collection area, which is
estimated at about 10 $%
\mu
$m in the spectral range of GaAs QW radiation due to chromatic aberration. The
microscopic sample image with the laser spot focused to the channel is shown
in Fig. 2(a)\textbf{.} The PL spatial resolution of the setup is determined by
the size of the light collection area, which is estimated at about 10 $\mu$m
due to chromatic aberration. Circularly polarized PL was measured in a
magnetic field of 9T applied perpendicular to the QW plane. Similar results
were obtained in different polarizations. Therefore, the data obtained in one
$\sigma^{-}$ polarization are presented below. A detailed description of the
sample and experimental setup can be found in \cite{pusep2025a,pusep2025b}.

\section{Results and Discussion}

The influence of the electric current flowing within the channel when a
potential difference is applied to the ends of the channel on the magneto-PL
of the hydrodynamic e-h plasma is demonstrated in Fig. 1. Without current in
the channel, PL lines corresponding to transitions between the electron and
heavy hole Landau levels are observed \cite{pusep2025b}. The best fit of the
calculated Landau fan to the experimental one was obtained with the exciton
effective mass 0.059m$_{0}$ which agrees well with the heavy hole exciton
effective mass 0.06m$_{0}$ \cite{exmass}. Weak PL line LH0 is found due to the
lowest light hole LL.
\begin{figure}
    \centering
\includegraphics[width=8 cm]{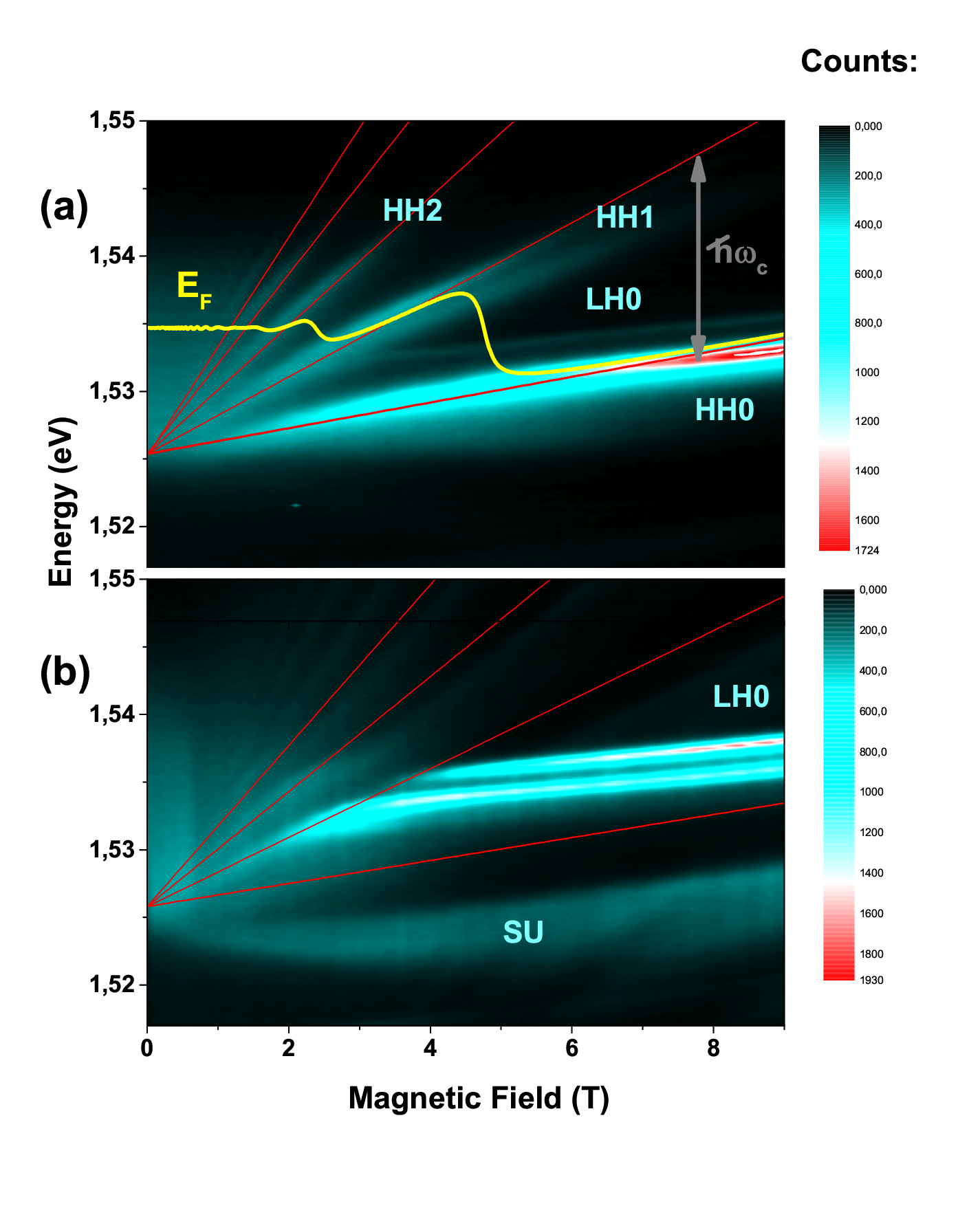}
\caption{(Color on line) Contour plot of PL measured from an unbiased channel
(a) and with a current of 100 $\mu$A parallel to the channel (b) as a function
of the magnetic field at T = 4K. The calculated Landau fan energies and Fermi
level energy are shown as red and yellow lines, respectively. HH and LH denote
the optical transitions caused by heavy and light holes, respectively. The
double PL line in (b) is due to a light hole exciton (higher emission energy)
and a positively charged light hole trion (lower emission energy). SU is the
PL line due to the shake-up process involving light holes. Data taken from
\cite{pusep2025b}.}
\end{figure}

As shown in Fig. 1(b), the electric current in the channel radically changes
the magneto-PL: instead of the Landau fan, a double PL line appears at the
critical current. The potential difference applied across the ends of the
channel gives the electron fluid a net drift velocity. As shown in
\cite{pusep2025b}, under the laser radiation the electron flow in the channel
causes the drag of photogenerated light holes, while the dynamics of heavy
holes remains virtually unchanged. In the region near the laser excitation
point, where photogenerated holes diffuse in the direction opposite to the
electron flow, electrons slow down the propagation of light holes. As a
result, in this region, light holes accumulate, and their increasing local
concentration creates conditions for the emergence of excitons (PL line with
higher energy) and exciton complexes in the form of positively charged trions
(PL line with lower energy). The emission labeled SU arises from a shake-up
process involving the lowest LLs of light holes, which was studied in detail
in \cite{pusep2025b}.

Another experimental configuration, shown in Fig. 2(a), allowed us to separate
the effects of electric current and electric field. In this case, no electric
field is applied to the ends of the channel and, therefore, no current flows
inside the channel. At the same time, the potential difference is applied to
the potentiometric contact pads, resulting in a corresponding Hall current
I$_{H}$ perpendicular to the channel. The magnetic field applied normal to the
QW plane causes a Lorentz force on the drifting background electrons,
resulting in Hall voltage difference across the channel.

\begin{figure}
    \centering
\includegraphics[width=8 cm]{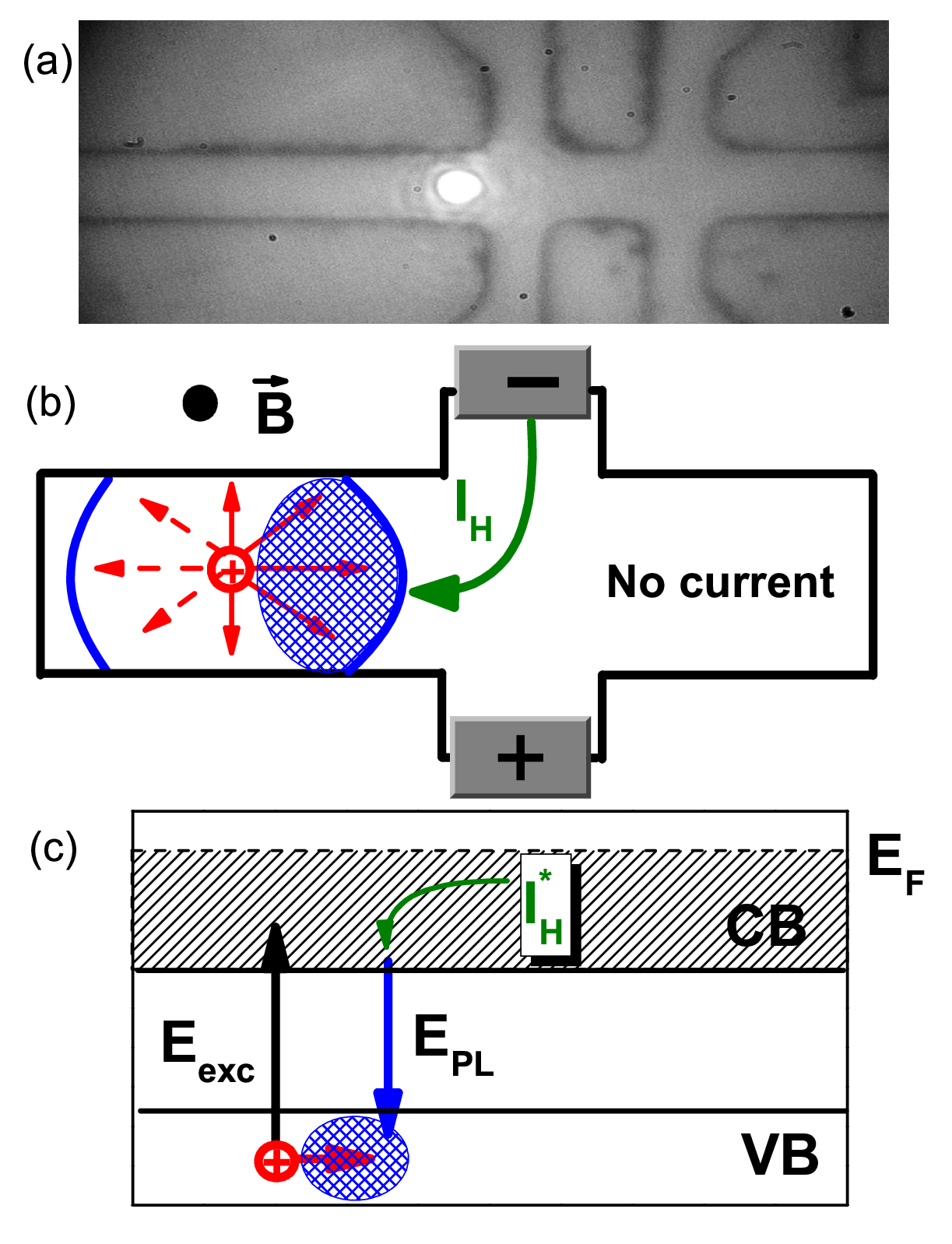}
\caption{(Color on line). The microscopic sample image with the laser spot
focused to the channel (a), experimental scheme (a) and energy diagram of the
sample (b). The drift trajectory of electrons in an applied magnetic field is
shown by green arrow and the boundaries of the PL collection area are
indicated by blue brackets in (a). The laser excitation point is shown as the
position of the positive charge of holes that are photogenerated in the QW and
injected from the barriers. Diffusion of the holes is shown by red arrows in
(a), and the region there they accumulate is indicated by the blue shaded
area. The effective electron current I$_{H}^{\ast}$ caused by the
recombination of electrons with photogenerated holes is shown by green arrow
in (b).}
\end{figure}
Under intense laser excitation with an energy of 2.82 eV exceeding the gap in
the barriers, the electric field built in the barriers spatially separates the
electrons and holes photogenerated in the barriers, leading to the injection
of holes into the QW, where they form hydrodynamic e-h plasma with nearly
equal concentrations of electrons and holes \cite{pusep2023} within a region
determined by the hole diffusion length (about 5-10 $\mu$m) \cite{pusep2025a}.
It is worth noting that, despite the strong magnetic field, in the region
where e-h plasma arises, the quantum Hall regime does not develop due to
intense e-h scattering.

The diffusion of photogenerated holes in the PL collection area, schematically
shown in Fig. 2(a), significantly changes the charge distribution in the
channel. Recombination occurring in the hole diffusion region leads to a rapid
change in the local electron population and thus opens a channel for an
effective interband current I$_{H}^{\ast}$, shown in Fig. 2(b)
\cite{ma2013,amol2024}. As a result, an effective Hall current flows in the
illuminated part of the channel, and a Hall voltage is applied to its dark
part. The effective Hall current affects hole diffusion similarly to the
channel current: in the PL collection area, where the concentration of
injected holes is high, Coulomb e-h drag leads to the selection of light and
heavy holes, since electrons drag light holes much more strongly than heavy
holes. As a result, an effective Hall current flows in the illuminated part of
the channel, and a Hall voltage is applied to its dark part. The effective
Hall current affects hole diffusion similarly to the channel current: in the
PL collection area, where the concentration of injected holes is high, Coulomb
e-h drag leads to the selection of light and heavy holes, since electrons drag
light holes much more strongly than heavy holes. Accordingly, the diffusion of
light holes is inhibited by the Hall current in the opposite direction, which
causes accumulation of light holes, shown in Fig. 2 by the blue shaded area.
Thus, Coulomb e-h drag leads to the accumulation of light holes in the
illuminated region, where they begin to dominate the PL emission.

It should be noted that by the Hall voltage we mean the potential difference
across the channel associated with the Hall effect, whereas by the Hall
current we mean the effective current in the illuminated region of the channel
arising due to the redistribution of charge density caused by the
recombination of photogenerated carriers.

The PL\ measured along the channel in a magnetic field of 9T and with a
current of 100 $\mu$A parallel to the channel is shown in Fig. 3(a). As shown
in \cite{pusep2025b}, an increase in the concentration of light holes in the
illuminated region creates conditions for the formation of an exciton $X_{LH}$
and a positively charged trion $X_{LH}^{+}$, which appear in Fig. 1(b) as a
double PL line. The data presented in Fig. 3(b) show that if instead a current
of 100 $\mu$A flows perpendicular to the channel, the effective Hall current
in the channel produces the same double PL line, consisting of excitons and
trions. While in the opposite section of the channel, where no current flows,
PL lines due to the LLs of heavy holes HH0 and HH1 are observed. The weak
dependence of the LL energy on the distance is due to the Hall voltage applied
to this section of the channel. Without a current in the channel, the expected
Hall voltage drop along the channel width is approximately 0.6 V. The
effective Hall current flowing in the channel sharply reduces the Hall
voltage. The change in LL energy along the channel corresponds to a small
residual fraction of the Hall voltage of 3 mV due to the small fraction of
electrons reaching the left contact pad in Fig. 2(a).
\begin{figure}
    \centering
\includegraphics[width=8 cm]{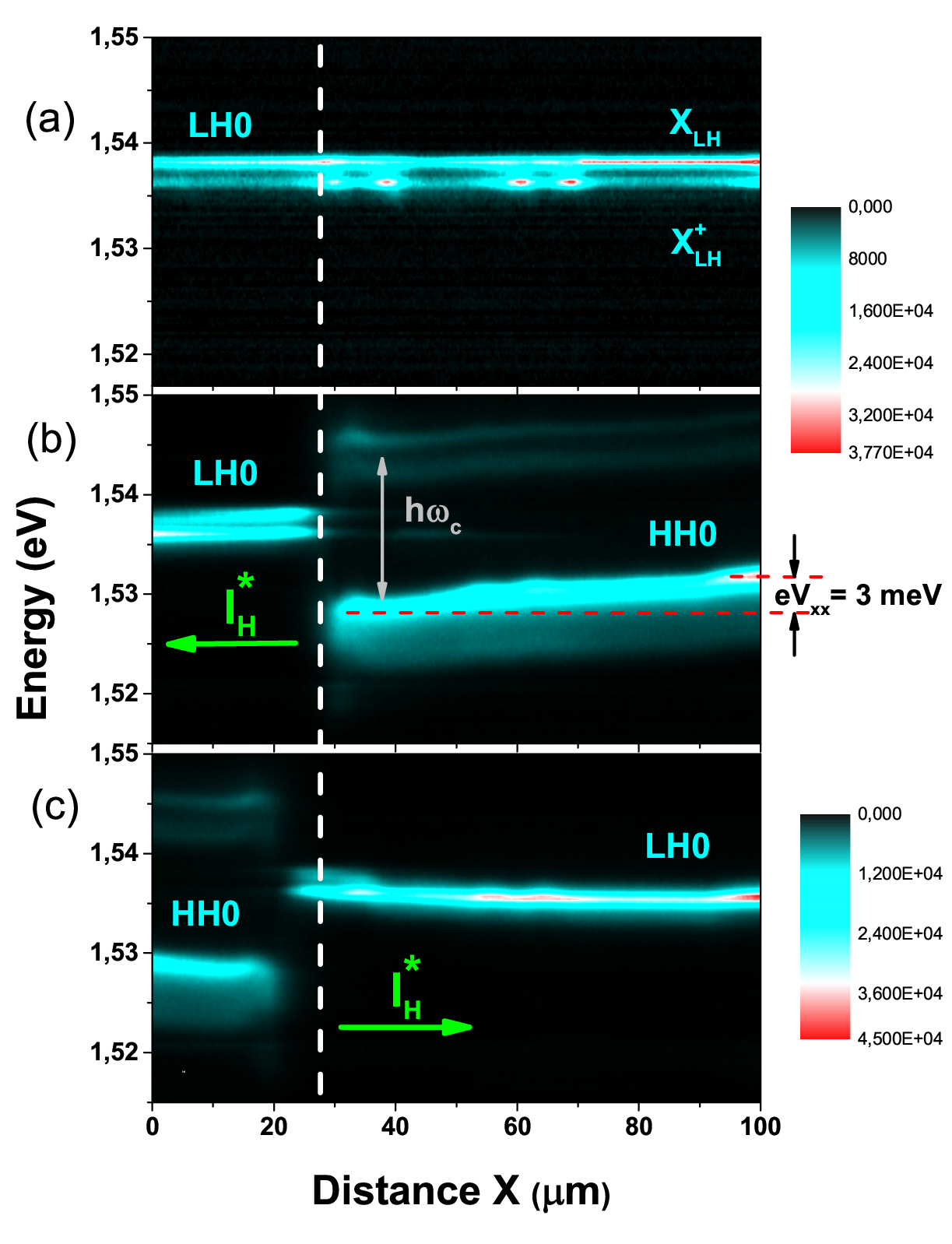}
\caption{(Color on line). Contour plot of PL measured at a temperature of 4K and
a magnetic field of 9T, as a function of the distance along the channel. With
a current of 100 $\mu$A parallel to the channel and unbiased potentiometric
contacts (a), with an unbiased channel and a perpendicular current of 100
$\mu$A through potentiometric contacts, with negative and positive potential
at the upper and lower contacts, respectively (b) and with reverse polarity
(c). The position of the potentiometric contacts is shown by the vertical
white dashed line. The direction of the effective electron Hall current
I$_{H}^{\ast}$ is shown by green arrows.}
\end{figure}

As expected and shown in Fig. 3(c), the nature of the PL in different sections
of the channel is reversed when the polarity of the electric current through
the potentiometric contacts is changed.

It is worth noting that the aforementioned phenomenon, caused by charge
accumulation under the influence of an electric current, is similar to the
processes occurring in light-emitting diodes (LEDs). The fundamental
difference between current-stimulated emission in an LED and the described
case is determined by the charge accumulation processes. An LED is a
combination of two types of semiconductor materials, p-type and n-type, in a
single crystal, which forms a potential difference across the p-n junction
where charge accumulation occurs in the depletion region in a forward-biased
LED. In this case, the increase in the electron and hole concentrations in the
depletion region leads to radiative recombination associated with the
interband current \cite{mane2023}. In contrast, in the device reported here,
electric-field-stimulated emission occurs in a uniformly n-doped QW, while e-h
drag leads to charge accumulation. In both cases, electron recombination leads
to the generation of an interband current.

\section{Conclusion}

In conclusion, we observed effect of the Hall current on the PL of a dense
hydrodynamic e-h plasma formed in a mesoscopic GaAs channel. The effective
Hall current, caused by interband recombination, leads to the accumulation of
photogenerated light holes due to Coulomb drag, and, consequently, to the
appearance of a double photoluminescence line consisting of excitons and
trions. Meanwhile, in the absence of a Hall current, the change in the PL
energy of heavy holes is caused by the electric field generated by the Hall
potential difference. In this way experimental evidence was obtained for the
existence of two different photoluminescence mechanisms in the same mesoscopic
GaAs system, caused by electric field effects and current-induced effects. The
observed effect demonstrates electrical control of exciton complex populations
through current-induced redistribution of charge carriers in a uniformly doped
n-type quantum well without a p-n junction, where the current geometry and
magnetic field can be used to control the local optical response. Moreover,
our study shows that the optical response can serve as a local probe of
hydrodynamic electron-hole transport, since the appearance of the
exciton/trion double line is associated with current-driven hole accumulation
mediated by strong electron-hole scattering.

\textbf{Acknowledgments:} Financial supports from the Brazilian agencies
FAPESP (Grants 2021/12470-8, 2022/10340-2) are gratefully acknowledged.

\end{document}